# Enhancing and suppressing radiation with some permeability-near-zero structures

Yi Jin[1,2] and Sailing He[1,2,3,*]

[1]*Centre for Optical and Electromagnetic Research, State Key Laboratory of Modern Optical Instrumentations, Zhejiang University, Hangzhou 310058, China*
[2]*Joint Research Laboratory of Optics of Zhejiang Normal University and Zhejiang University, China*
[3]*Division of Electromagnetic Engineering, School of Electrical Engineering, Royal Institute of Technology, S-100 44 Stockholm, Sweden*
*sailing@kth.se*

**Abstract:** Using some special properties of a permeability-near-zero material, the radiation of a line current is greatly enhanced by choosing appropriately the dimension of a dielectric domain in which the source lies and that of a permeability-near-zero shell. The radiation of the source can also be completely suppressed by adding appropriately another dielectric domain or an arbitrary perfect electric conductor (PEC) inside the shell. Enhanced directive radiation is also demonstrated by adding a PEC substrate.

## References and links

## 1. Introduction

Metamaterials can achieve arbitrary optical constants in principles, which provide us flexible manipulation of the behavior of electromagnetic waves. Based on the theoretical work of Pendry et al. [1,2], it was experimentally demonstrated at microwave frequencies by Smith et al. [3] that an artificial microstructured composite could exhibit a negative refractive index. Later, superlens and cloaks were also suggested, and experimentally demonstrated [4−7]. Among various unusual material parameters at microwave and optical frequencies provided by metamaterials, near zero permittivity/permeability is a singular material parameter. Polar materials and noble metals can also provide near zero permittivity near their resonant frequencies at infrared and visible frequencies, respectively. Near zero permittivity/permeability can lead to many interesting phenomena and applications [8−17]. Any electromagnetic wave incident on such a material is totally reflected except nearly normal incidence. Based on this property, such a material can be used to enhance radiation directivity and spatial filtering [9−11]. It was also suggested that near zero permittivity/permeability can be used for squeezing electromagnetic energy [12−15], shaping the phase front [16], and transmitting subwavelength image [17]. In this paper, we will show that when a common dielectric domain with an internal line current is surrounded with a permeability-near-zero shell, the radiation of the source can be strongly enhanced. The radiation can also be completely suppressed by simply adding another common dielectric domain or a perfect electric conductor (PEC) inside the shell. Enhanced directive radiation based on the permeability-near-zero shell with a PEC substrate is demonstrated.

## 2. Enhancing radiation

In the present study, two-dimensional electromagnetic propagation is assumed with the electric field perpendicular to the *x-y* plane, and the time harmonic factor is $exp(-i\omega t)$. The structure shown in Fig. 1(a) is investigated first. Domain 4 is not considered (i.e., domain 4 vanishes) for the moment. Domain 3 is surrounded by domain 2 of relative permeability $\mu_2$ as the shell, and domain 1 is the background. Domains 1 and 3 are of common dielectric. A unit line current propagating along the *z* axis is inside domain 3 as an excitation source. Let $F_n(\mathbf{r})$ denote some quantity $F(\mathbf{r})$ in domain $n$ ($n=1,2,3$). In domain 2, the electromagnetic field satisfies Maxwell's equations,

$$\frac{\partial E_{2,z}(\mathbf{r})}{\partial y}\mathbf{e}_x - \frac{\partial E_{2,z}(\mathbf{r})}{\partial x}\mathbf{e}_y = i\omega\mu_2\mu_0[H_{2,x}(\mathbf{r})\mathbf{e}_x + H_{2,y}(\mathbf{r})\mathbf{e}_y], \tag{1}$$

$$\frac{\partial H_{2,y}(\mathbf{r})}{\partial x} - \frac{\partial H_{2,x}(\mathbf{r})}{\partial y} = -i\omega\varepsilon_0 E_{2,z}(\mathbf{r}), \tag{2}$$

where $\varepsilon_0$ and $\mu_0$ are the vacuum permittivity and permeability, respectively, and subscripts *x*, *y*, and *z* represent a component of an electromagnetic quantity along some canonical base. One can rewrite Eq. (2) for the shell region in the following integration form,

$$\oint_{\partial 2}\mathbf{H}_{2,\partial 2}(\mathbf{r})\cdot d\mathbf{l} + \oint_{\partial 3}\mathbf{H}_{2,\partial 3}(\mathbf{r})\cdot d\mathbf{l} = -i\omega\varepsilon_0\int_2 E_{2,z}(\mathbf{r})ds, \tag{3}$$

where integration contours $\partial 2$ and $\partial 3$ are shown in Fig. 1(a), and $\mathbf{H}_{2,\partial n}(\mathbf{r})$ denotes the value of magnetic field $\mathbf{H}_2(\mathbf{r})$ at boundary $\partial n$. When $\mu_2$ tends to zero, electric field $E_{2,z}(\mathbf{r})$ in domain 2 is uniform and assumed to be constant $E_z$. Then, the right side of Eq. (3) can be written as $-i\omega\varepsilon_0 S_2 E_z$ where $S_2$ is the area of domain 2. According to the continuity condition of the tangential electric and magnetic fields at both sides of a boundary [18], $E_{1,z,\partial 2}(\mathbf{r})$ and $E_{3,z,\partial 3}(\mathbf{r})$ are also equal to $E_z$, and the contour integrals of $\mathbf{H}_{2,\partial n}(\mathbf{r})$ at the left side of Eq. (2) can be replaced by those of $\mathbf{H}_{1,\partial 2}(\mathbf{r})$ and $\mathbf{H}_{3,\partial 3}(\mathbf{r})$, respectively. Following the unique theorem [18], if one knows the tangential electric field at the boundary of a domain and the excitation source in the domain, the electromagnetic field in the whole domain is uniquely determined. Therefore, both sides of Eq. (3), as well as the fields in domains 1 and 3, are independent of the position of domain 3 inside domain 2.

If both domains 2 and 3 are circular with radii $r_2$ and $r_3$, respectively, and the unit line current is at the center of domain 3, the value of $E_z$ can be analytically obtained. The electromagnetic field in each domain can be expanded by the Bessel and Hankel functions [18]. Since the electric field on boundary $\partial 3$ is equal to constant $E_z$ everywhere, and the line current is at the center of domain 3, the electric field in domain 3 is only composed of the zero-order Bessel and Hankel functions,

$$E_{3,z}(\mathbf{r}) = H_0(k_3 r) + a_3 J_0(k_3 r), \tag{4}$$

where $k_3$ is the wave-number in domain 3. Similarly, the electric field in domain 1 is only composed of the zero-order Bessel function,

$$E_{1,z}(\mathbf{r}) = a_1 H_0(k_1 r). \tag{5}$$

In Eqs. (4) and (5), the polar coordinate origins locate at the centers of domains 3 and 2, respectively. In Eqs. (4) and (5), unknown coefficients $a_1$ and $a_3$ can be expressed as functions of $E_z$ based on the boundary condition that both $E_{1,z,\partial 2}(\mathbf{r})$ and $E_{3,z,\partial 3}(\mathbf{r})$ are equal to $E_z$. When electric field $E_{n,z}(\mathbf{r})$ in domain $n$ is known, the corresponding magnetic field $\mathbf{H}_n(\mathbf{r})$ can be obtained from Maxwell's equations. Then, according to Eq. (3), one obtains the value of $E_z$,

$$E_z = k_3 r_3 [H_1(k_3 r_3) - \frac{J_1(k_3 r_3) H_0(k_3 r_3)}{J_0(k_3 r_3)}] / [\frac{k_1 r_2 H_1(k_1 r_2)}{H_0(k_1 r_2)} - \frac{k_3 r_3 J_1(k_3 r_3)}{J_0(k_3 r_3)} - \frac{\omega^2 \varepsilon_0 \mu_0 S_2}{2}]. \tag{6}$$

The radiated power of the source into the background is most interesting. The ratio between the radiated power in the present case and that when the same source is positioned in infinite domain 1 is defined as power enhancement factor $P_{norm}$, i.e.,

$$P_{norm} = |a_1|^2 = |\frac{E_z}{H_0(k_1 r_2)}|^2. \tag{7}$$

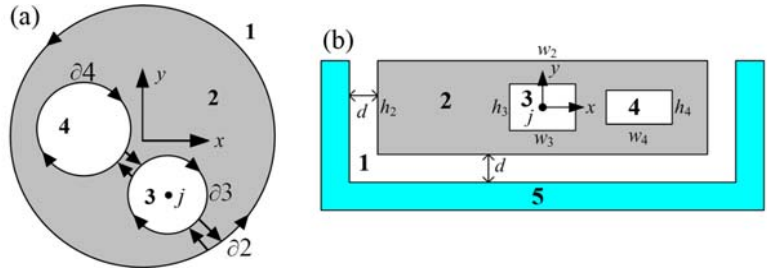

Fig. 1. (a) Configuration of domain 2 with internal domains 3 and 4. Domain 1 is the background. (b) Configuration when domains 2−4 in (a) are replaced with rectangular ones and domain 5 is added in the background.

Now we study the denominator of Eq. (6) (denoted as $F$). If $|F|$ can be small, $|E_z|$ and $P_{norm}$ will be large. There are three items in the expression of $F$ whose real parts can cancel each other. When domains 1 and 3 are lossless, since there is only one complex item in the expression of $F$, imag($F$) can not be zero, unlike real($F$). However, if |imag($F$)| can be small, large $|E_z|$ and $P_{norm}$ can still be obtained. This condition can be fulfilled as shown in the following example. Domains 1 and 3 are assumed to be of free space. When $r_2 = 2\lambda_0$ ($\lambda_0$ is the wavelength in free space), Fig. 2 shows the real and imaginary parts of $F$ as well as the corresponding $P_{norm}$ as $r_3$ varies. As shown in Fig. 2(a), real($F$) is zero at some special values of $r_3$. The appearance of several zero roots of real($F$) is because the Bessel functions are oscillating functions. As shown in Fig. 2(b), $P_{norm}$ is locally maximal at these special values of $r_3$, and the radiation is greatly enhanced. When $r_3 = 0.38775\lambda_0$, $P_{norm}$ is about 190. The whole structure behaves as a resonator to enhance the radiation. Because real($F$) steeply passes through the zero points as shown in Fig. 2(a), the peaks of $P_{norm}$ in Fig. 2(b) are very narrow. This indicates that the enhanced radiation is sensitive to the dimension of domain 3. In the inset of Fig. 2(b), one can see that to keep $P_{norm}$ in the same order of magnitude and strongly enhance the radiation, three figures are necessary for $r_3$. The radiation can also be influenced by the dimension of domain 2, which can be stronger enhanced as $r_2$ increases. This may be in contrast to our common intuition. Based on the fact that the normal transmission through a permeability-zero slab is approximately inversely proportional to the thickness [10], one may think that the radiation will be weaker as domain 2 becomes larger.

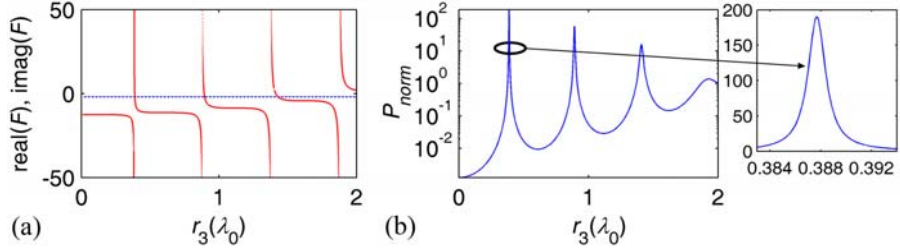

Fig. 2. (a) real($F$) and imag($F$). The red (dotted) and blue (dashed) curves correspond to real($F$) and imag($F$), respectively. (b) $P_{norm}$ as a function of $r_3$ with $r_2=2\lambda_0$.

## 3. Completely suppressing radiation

As shown in Fig. 2(b), if the dimension of domain 3 is not chosen appropriately, the radiation of the line current is suppressed instead of enhanced, but not completely. In this section, we will completely suppress the radiation by adding a common dielectric domain or a perfect electric conductor.

First we add a common dielectric domain (as domain 4) of radius $r_4$ inside domain 2. Fig. 3 shows $P_{norm}$ as a function of radius $r_4$ when domain 4 is circular of free space, $r_2=2\lambda_0$, and $r_3=0.38775\lambda_0$. The existence of domain 4 greatly influences the radiation. When domain 4 is large, $P_{norm}$ is very large only in a narrow range of $r_4$. When $r_4$ is at some special values, $P_{norm}$ is zero. This can be understood in the following way. The electric field inside domain 4 is just composed of the zero-order Bessel function, $a_4 J_0(k_4 r_4)$, and $E_z$ can still be expressed by Eq. (6) after adding in the denominator a term of $-k_4 r_4 J_1(k_4 r_4)/J_0(k_4 r_4)$. $J_0(x)$ is an oscillating function with zero roots. When $k_4 r_4$ is a zero root of $J_0(x)$, electric field $E_{4,z,\partial 4}(\mathbf{r})$ is zero on boundary $\partial 4$, and consequently $E_z$ and $E_{1,z,\partial 2}(\mathbf{r})$ are forced to be zero. This means that no power can be radiated into the background, i.e., the radiation is completely suppressed. This is a rather interesting result. Just by adding a common dielectric material, permeability-zero domain 2 behaves like a PEC, inside which the electric field is zero. If the excitation source is moved outside domain 2, the radiated wave seems scattered by a PEC possessing the shape of domain 2. However, domain 2 is not completely like a PEC, as the internal magnetic field is not zero, which will be shown below.

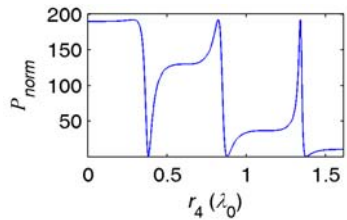

Fig. 3. $P_{norm}$ as a function of $r_4$ with $r_2=2\lambda_0$ and $r_3=0.38775\lambda_0$.

Besides the above approach, we can take another approach to completely suppress the radiation of the line current. It is based on the fact that the electromagnetic field is zero in a

PEC and the electric field in a permeability-zero material is uniform. When a PEC crosses or locates inside domain 2, their boundary condition forces the electric field in whole domain 2 to be zero. The PEC can be of arbitrary shape and dimension. This leads to no power radiated into the background.

## 4. Numerical simulations

To enhance and suppress the radiation, the domains in Fig. 1(a) can be of other shapes besides circular shapes. Here, rectangular domains are used as examples to enhance or suppress the radiation. When domains 1−4 in Fig. 1(a) are replaced by corresponding rectangular domains and domain 5 (PEC) is added in the background, the structure is shown in Fig. 1(b). Unlike in a circular domain, the field distributions in rectangular domains 2, 3 and 4 can not be expanded by only several simple base functions. Thus, we perform the numerical simulation with a finite-element-method software, COMSOL [19]. In the simulation, domains 1, 3, and 4 are of free space.

We first study the case when only domains 1−3 exist in Fig. 1(b). The permeability of domain 2 is assumed to be $\mu_2=10^{-5}+10^{-4}i$, which deviates a bit from zero. The width and height of domain 2 are $w_2=8\lambda_0$ and $h_2=\lambda_0$, respectively. The centers of domains 2 and 3 are at the same point with the line current. When the width and height of domain 3 are appropriately chosen, the radiation of the line current can be strongly enhanced. Such an example is $w_3=0.8\lambda_0$ and $h_3=0.66\lambda_0$, for which $P_{norm}$ is 46. Fig. 4(a) shows a snapshot of the electric field distribution and Fig. 4(b) shows the magnetic amplitude distribution. The electromagnetic wave is radiated in all the directions, and most energy flows nearly normally through the long top and bottom surfaces of domain 2.

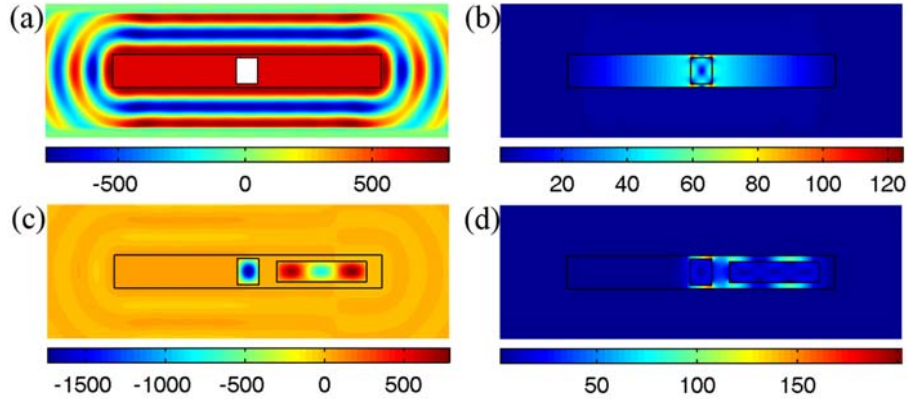

Fig. 4. Electromagnetic field distributions of the structure in Fig. 1(b). (a) and (c) are the snapshots of the electric field distributions, and (b) and (d) are the magnetic amplitude distributions. In (a) and (b) only domains 1−3 exist, and in (c) and (d) domains 1−4 exist. In (a), the field distribution in the white box is removed in order to show clearly those in the other domains.

Then rectangular domain 4 is added inside domain 2 to suppress the radiation. If the radiation is completely suppressed and the electric field in domain 2 is zero, the electromagnetic field will oscillates like in a PEC cavity, in which the electric component should be even in symmetry since the electric field in domain 2 is uniform. According to the existence condition of the modes in a rectangular PEC cavity, to completely compress the radiation, the dimension of domain 4 should fulfill

$$\begin{aligned} w_4 &= m\pi/k_x, \quad m=1,3,5... \\ h_4 &= n\pi/k_y, \quad n=1,3,5... \\ k_x^2 + k_y^2 &= (2\pi/\lambda_0)^2 \end{aligned} \qquad (8)$$

As an example, it is assumed that $k_x=3k_y/2$, $m=1$, and $n=3$, then one can get the corresponding values of $w_4$ and $h_4$ according to Eq. (8). Fig. 4(c) shows a snapshot of the electric field distribution, and Fig. 4(d) shows the magnetic amplitude distribution. The electric field in domains 1 and 2 is very weak, but not exactly zero because $\mu_2$ deviates a bit from zero, and the radiation is also very weak with $P_{norm}$ only 0.1. The magnetic field in domain 2 is nonzero and rather non-uniform.

Now the influence of PEC domain 5 on the radiation is investigated with domain 4 removed. If domain 5 is very close to domain 2, it can be considered as a transition to the limit case when domain 5 touches domain 2, the radiation can still be suppressed. The magnetic field in the narrow gap between domains 2 and 5 is very strong. If domain 5 is moved away from domain 2, the suppression of the radiation is gradually weakened. In fact, as shown below, on the contrary the existence of domain 5 may even help enhance the radiation.

In directive radiation, it is usually required that the electromagnetic field propagates only along one direction. From Figs. 4(a) and 4(b) one sees that the electromagnetic field mainly propagates through both the top and bottom surfaces. To realize directive radiation, a PEC substrate possessing the shape of domain 5 in Fig. 1(b) is used. Fig. 5(a) shows $P_{norm}$ as a function of the distance ($d$) between domains 2 and 5. When $d$ is small, $P_{norm}$ is small (as expected). When $d$ increases gradually, $P_{norm}$ also increases. When $d=0.38\lambda_0$, the radiation of the line current is most strongly enhanced, and $P_{norm}$ reaches 111, which is larger than that when domain 5 is not added. This indicates that when $d$ is appropriately chosen, the reflected wave by domain 5 can coherently help enhance the radiation. Fig. 5(b) shows a snapshot of the electric field distribution, and Fig. 5(c) shows the magnetic amplitude distribution, in which one can see clearly that the upward propagation of the radiated electromagnetic field is very directive.

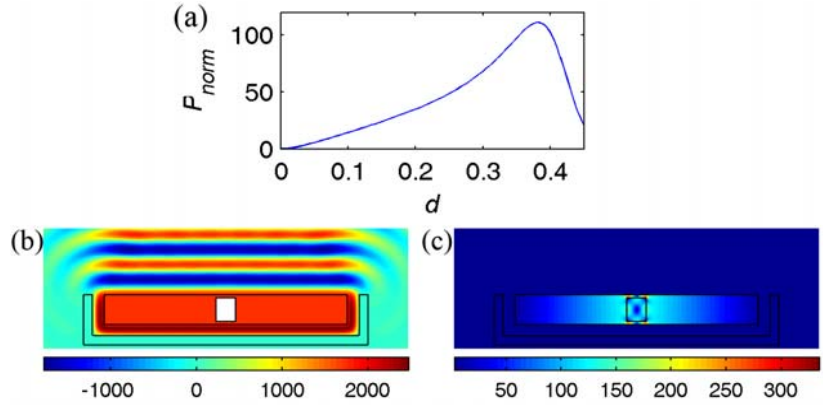

Fig. 5. Radiated power and electromagnetic field distributions for the structure of Fig. 1(b) when domain 5 is added and domain 4 is removed. (a) $P_{norm}$ as $d$ varies, (b) snapshot of the electric field distribution, and (c) the magnetic amplitude distribution.

## 5. Conclusion

We have shown the radiation of a line current surrounded by a permeability-near-zero shell can be greatly enhanced or completely suppressed in different circumstances. We have also demonstrated a structure of enhanced directive radiation. Similarly, according to the reciprocity of Maxwell's equations, a permittivity-near-zero shell should have similar results. The results can also be generalized to enhancement and suppression of radiation in three-dimensional space.

## Acknowledgments

This work is partially supported by the National Natural Science Foundation (Nos. 60990320, 60901039) of China.